\documentclass[a4paper,12pt]{article}

\usepackage[left=2.5cm,right=2.5cm,top=2.5cm,bottom=2.5cm]{geometry}
\usepackage{graphicx}
\usepackage{amssymb}
\usepackage{multirow}
\usepackage{amsmath}
\usepackage{psfrag}
\usepackage{setspace}
\usepackage{subcaption}
\usepackage[colorlinks=true,bookmarks=true]{hyperref}
\usepackage{float}
\usepackage{cite}
\usepackage{authblk}

\usepackage{mathtools}

\DeclarePairedDelimiterX\braket[2]{\langle}{\rangle}{#1 \delimsize\vert #2}

\hypersetup{citecolor=blue}
\newcommand{\dif}{\mathrm{d}}
\newcommand{\Eqref}[1]{(\ref{#1})}
\newcommand{\half}{\frac{1}{2}}
\newcommand{\expo}[1]{\mathrm{e}^{#1}}
\newcommand{\brac}[1]{\left(#1 \right)}
\newcommand{\sbrac}[1]{\left[#1\right]}
\newcommand{\im}{\mathrm{i}}

\begin{document}

\title{Kasner-type singularities and solitons with Lifshitz asymptotics}
\author{Yen-Kheng Lim\thanks{Email: phylyk@nus.edu.sg}}
\affil{\normalsize{\textit{Department of Physics, National University of Singapore, 117551, Singapore}}}

\renewcommand\Authands{ and }

\date{\normalsize{\today}}
\maketitle

\begin{abstract}
  We present an exact solution in Einstein-Maxwell-dilaton gravity describing a spacetime with an anisotropic Kasner-type singularity and Lifshitz asymptotics. This configuration can also be supported by a phantom scalar while still satisfying the Null Energy Condition. For certain parameters of this solution, null geodesics can have an infinitely deep effective potential, thus trapping photons in a finite region along the radial direction. Some examples of periodic null geodesics are obtained. A particularly interesting special case of this solution is a regular, soliton-type metric that retains its Lifshitz scaling in the time coordinate.
\end{abstract}

\section{Introduction} \label{intro}

The holographic correspondence provides a map between a $d$-dimensional strongly coupled field theory to a $(d+1)$-dimensional gravity with a negative cosmological constant. One of the earliest examples of this is the well-known AdS/CFT correspondence, which maps a $d$-dimensional conformal field theory (CFT) to a low energy limit of string theory, which is an Einstein gravity with a negative cosmological constant described by the Anti-de Sitter (AdS) spacetime in $d+1$ dimensions. 

Since Maldacena's original conjecture, there has been the many proposed applications of this correspondence. One recent area of interest is to study the holographic duals of non-relativistic field theories. In particular, asymptotically Lifshitz spacetimes are dual to condensed matter systems with Lifshitz fixed points \cite{Kachru:2008yh,Balasubramanian:2009rx}. (See also Refs.~\cite{Taylor:2008tg,Taylor:2015glc} and references therein.) Such systems are considered non-relativistic as they scale anisotropically between the space and time directions as
\begin{align}
 t\rightarrow\lambda^z t,\quad \vec{x}\rightarrow\lambda\vec{x},
\end{align}
for constants $\lambda$ and $z$, where the latter is called the \emph{Lifshitz exponent}. There have since been many studies on various aspects of asymptotically Lifshitz spacetimes and their possible applications to their non-relativistic duals \cite{Balasubramanian:2009rx,Sin:2009wi,Pang:2009wa,Sun:2013zga,Foster:2016abe}. 

Certain systems of interest also have anisotropies among the spatial directions as well. For instance, Ref.~\cite{Cremonini:2014pca} considered a system which requires a spatially anisotropic configuration. Kachru et al. \cite{Kachru:2013voa} considered the possible geometries that allow a holographic renormalisation-group flow between an asymptotic Lifshitz spacetime and a spatially anisotropic region. This anisotropic region is described by Bianchi-attractor geometries of various types \cite{Iizuka:2012iv,Iizuka:2012wt}. Using the notation in our present paper, the holographic flow is parametrised by the Poincar\'{e}-type coordinate $0<u<u_0$, where $u\rightarrow 0$ refers to the boundary field theory in the UV regime, and $u=u_0$ is typically referred to as the deep-IR region of the bulk. In \cite{Roychowdhury:2015cva}, approximate and numerical solutions of spatially anisotropic solutions with Lifshitz asymptotics have found.

A particular metric in the Bianchi Type I class is the Kasner spacetime. In its original form, the Kasner metric is a Ricci-flat cosmological solution with a singularity at the initial time $t=0$. This solution also has a `radial' version, where in $D=(d+1)$-dimensions is given by
\begin{align}
 \dif s^2&=-u^{2\beta_0}\dif t^2+\sum_{i=1}^{D-2}u^{2\beta_i}\brac{\dif x^i}^2+\dif u^2,
\end{align}
where the exponents are required to satisfy the \emph{Kasner conditions}
\begin{subequations}\label{KasnerConditionsOri}
\begin{align}
 \beta_0+\sum_{i=1}^{D-2}\beta_i&=1, \label{KasnerConditionsOri1}\\
 \beta_0^2+\sum_{i=1}^{D-2}\beta_i^2&=1.\label{KasnerConditionsOri2}
\end{align}
\end{subequations}
In the radial version, the singularity is time-like and is located at $u=0$. The Kasner, and more general singularities have been studied by Belinskii, Khalatnikov, and Lifshitz (and is thus known as BKL singularities) \cite{Belinsky:1970ew}. These singularities were recently studied in the context of the AdS/CFT correspondence in \cite{Engelhardt:2013jda,Engelhardt:2014mea,Shaghoulian:2016umj}.

Recently, Ren \cite{Ren:2016xhb} has obtained an exact solution where the metric is asymptotic to AdS in the UV regime and has a (radial) Kasner geometry in the IR limit. By varying the exponents in Ren's AdS-Kasner solution, one can interpolate between an AdS black hole and an AdS soliton \cite{Horowitz:1998ha}. In this paper, we shall present a generalisation of Ren's solution to include Lifshitz asymptotics. The solution derived below contains, as special cases, the Lifshitz black hole \cite{Kachru:2008yh,Taylor:2008tg,Pang:2009ad,Taylor:2015glc}, the AdS-scalar naked singularity \cite{Das:2001rk,Saenz:2012ga,Lim:2017dqw}, as well as Ren's AdS-Kasner solution. There is also a special case which we will interpret as the Lifshitz soliton. This is a different version from the Lifshitz soliton considered by Lu et al. \cite{Lu:2013tza}, which was constructed by performing a double-Wick rotation on a Lifshitz black hole. The soliton we obtain here retains its Lifshitz asymptotics in the time coordinate, as no Wick rotation is performed that turns it into a spatial one. In this sense, our Lifshitz soliton falls within the class of ansatz considered by \cite{Way:2012gr} to obtain soliton solutions numerically.\footnote{There is a different set of Lifshitz solitons, also known as \emph{Lifshitz stars}, which are localised objects with non-singular geometry and Lifshitz asymptotics, and are typically found as numerical solutions \cite{Danielsson:2009gi,Mann:2011bt}.} 
Similar to the Lifshitz soliton in \cite{Lu:2013tza}, one has to impose a periodicity on one of the spatial coordinates to remove a conical singularity.

The Kasner-like solution presented in this paper modifies one of the Kasner conditions, namely Eq.~\Eqref{KasnerConditionsOri2} where the right-hand side is replaced by free parameters related to the gauge and scalar field strengths. While the addition of matter to modify the Kasner conditions has been considered before in various contexts \cite{Koyama:2001rf,Das:2006dz,Awad:2007fj,Chatterjee:2016bhj}, the interesting feature of our solution is that the combined presence of the gauge and scalar fields allow the right-hand side to be restored to $1$. Additionally, we find a related result where the configuration also allows the existence of a phantom scalar field where the overall stress-energy tensor does not violate the Null Energy Condition.

It is well-known that only radial null geodesics are able to reach the boundary of Lifshitz spacetime\cite{Keeler:2013msa}, leading to various subtleties in the treatment of holographic quantities and must be dealt with care \cite{Gentle:2015cfp}. In any case, this is a consequence of the spacetime being dual to non-relativistic boundary by design, so that the speed of light is effectively infinite there. In the language of effective potentials of the geodesics, the photon encounters an infinite barrier at $u\rightarrow 0$. In the solution presented in this paper, we find that for certain choices of the Kasner exponents, the photon also encounters another infinite barrier at the IR singularity. This also occurs for the case of the regular Lifshitz soliton. In other words, classical null particles with non-zero transverse momentum in these spacetimes are trapped in an infinitely deep potential well. 

The rest of this paper is organised as follows: The metric and matter fields corresponding to our solution is presented in Sec.~\ref{Solution}. The Null Energy Condition for the solution is investigated in Sec.~\ref{NEC}, followed by Sec.~\ref{Geodesics} where we obtain the geodesic equations of motion for test particles in the spacetime. The specific case of the Lifshitz solitons is considered in Sec.~\ref{LifSol}. The parameter ranges of the solution for the four-dimensional case is studied in Sec.~\ref{Parameters}. A brief discussion and conclusion is given in Sec.~\ref{Conclusion}. The derivation of the solution is given in Appendix \ref{Derivation}.

\section{Equations of motion and the solution} \label{Solution}

Our solution is obtained under Einstein-Maxwell-dilaton gravity with a (negative) cosmological constant $\Lambda$ and an additional massless scalar $\psi$. We will show in Sec.~\ref{NEC} how $\psi$ is allowed to be a phantom scalar with the opposite sign of the kinetic term in the Lagrangian. In any case, for the present section let us write the Lagrangian for $\psi$ with the usual sign for the kinetic term.\footnote{Alternatively, one can write the Lagrangian for $\psi$ as $\mathcal{L}=\varepsilon\brac{\nabla\psi}^2$ and having $\varepsilon=\pm1$, but we prefer not to clutter the notation here.} As such, the action for this model is
\begin{align}
 I&=\frac{1}{16\pi G}\int\dif^Dx\sqrt{-g}\brac{R-2\Lambda-\expo{-2\alpha\varphi}F^2-\brac{\nabla\varphi}^2-\brac{\nabla\psi}^2}, \label{action}
\end{align}
where $F=\dif A$ is the Maxwell 2-form arising form a gauge potential $A$, and $\varphi$ is the scalar dilaton coupled to the Maxwell term via the coupling parameter $\alpha$. While both $\varphi$ and $\psi$ are clearly massless scalars in this model, for expositional convenience we shall henceforth refer to $\varphi$ as the \emph{dilaton}, and $\psi$ as the \emph{scalar} to distinguish between the two.

Let us briefly motivate the choice of the action given in Eq.~\Eqref{action}. The AdS-Kasner solution provided by Ref.~\cite{Ren:2016xhb} is a solution to pure Einstein gravity with a cosmological constant. As mentioned in Sec.~\ref{intro}, the inclusion of a massless scalar was considered in \cite{Koyama:2001rf,Das:2006dz,Awad:2007fj,Chatterjee:2016bhj}, which modifies one of the Kasner conditions. With the massless scalar, the action is $I=\frac{1}{16\pi G}\int\dif^Dx\sqrt{-g}\brac{R-2\Lambda-\brac{\nabla\psi}^2}$. Motivated by the holographic correspondence, exact and numerical solutions to this action have been explored in \cite{Das:2001rk,Saenz:2012ga}, where they tend to describe spacetimes with naked singularities. If an appropriate potential for the scalar is included, one may obtain black holes with AdS asymptotics \cite{Gonzalez:2013aca,Lu:2014maa}. If the kinetic term of the scalar carries an opposite sign, the action corresponds to a model with a phantom scalar with the action $I=\frac{1}{16\pi G}\int\dif^Dx\sqrt{-g}\big(R-2\Lambda+\big(\nabla\tilde{\psi}\big)^2\big)$. Notable solutions to this action include regular black holes \cite{Bronnikov:2005gm}. If a potential for $\tilde{\psi}$ is included, its cosmological solutions are well-known candidates to resolve the dark energy problem \cite{Caldwell:1999ew}. This phantom model have also been explored in the context of holography where its dual describe a high-temperature superconductor \cite{Lin:2015kjk}. With the rising interest of non-relativistic holography, it is desirable to generalise these results to Lifshitz asymptotics. In order to support such spacetimes, further matter fields are required. In this paper, we shall consider a dilaton-Maxwell field.\footnote{Alternatively, an Einstein-Maxwell gravity coupled to a massive vector field may also support Lifshitz spacetimes \cite{Kachru:2008yh,Taylor:2008tg}.} Taken together, we have the action as given in Eq.~\Eqref{action}.

Extremising the action gives the Einstein-Maxwell-dilaton-scalar equations
\begin{subequations}\label{EOM}
\begin{align}
 R_{\mu\nu}=\frac{2\Lambda}{D-2}g_{\mu\nu}&+2\expo{-2\alpha\varphi}F_{\mu\lambda}{F_\nu}^\lambda-\frac{1}{D-2}\expo{-2\alpha\varphi}F^2g_{\mu\nu}+\nabla_\mu\varphi\nabla_\nu\varphi+\nabla_\mu\psi\nabla_\nu\psi,\\
 \nabla_\lambda\brac{\expo{-2\alpha\varphi}F^{\lambda\nu}}&=0,\\
 \nabla^2\varphi+\alpha\expo{-2\alpha\varphi}F^2&=0,\\
 \nabla^2\psi &=0.
\end{align}
\end{subequations}
In component form, the Maxwell field is given by $F_{\mu\nu}=\nabla_\mu A_\nu-\nabla_\nu A_\mu$ and we denote $F^2=F_{\mu\nu}F^{\mu\nu}$.

The solution to be explored in this paper is given by
\begin{align}
 \dif s^2&=\ell^2\brac{-\frac{f^{\beta_0}\dif t^2}{u^{2z }}+\frac{\dif u^2}{u^2f}+\frac{1}{u^2}\sum_{i=1}^{D-2}f^{\beta_i}\brac{\dif x^i}^2},\quad f=1-\brac{\frac{u}{u_0}}^{z+D-2},\nonumber\\
 A&=\sqrt{\frac{\ell^2(z -1)}{2(z +D-2)}}\,u^{-(z +D-2)}\;\dif t,\nonumber\\
 \varphi&=-\sqrt{(z -1)(D-2)}\ln u+\frac{1-\beta_0}{2}\sqrt{\frac{z -1}{D-2}}\ln f,\quad \psi=-\frac{\eta}{2}\ln f,\nonumber\\
 \ell^2&=-\frac{(z +D-2)(z +D-3)}{2\Lambda},\quad\alpha=\sqrt{\frac{D-2}{z -1}}. \label{LifKas_Solution}
\end{align}
We show how this solution is derived in Appendix \ref{Derivation}. In order to solve the field equations in \Eqref{EOM}, the exponents $(\beta_0,\beta_1,\ldots,\beta_{D-2})$, are required to satisfy a modified version of Eq.~\Eqref{KasnerConditionsOri},
\begin{subequations}\label{KasnerConditions}
\begin{align}
 \beta_0+\sum_{i=1}^{D-2}\beta_i&=1,\label{KasnerCondition1}\\
 \beta_0^2+\sum_{i=1}^{D-2}\beta_i^2&=1-\frac{(z -1)(1-\beta_0)^2}{D-2}-\eta^2, \label{KasnerCondition2}
\end{align}
\end{subequations}
thus modifying the second Kasner condition by the presence of $z$, $\beta_0$, and $\eta$. Thus, we see that the constants $z $, $\eta$, and $(D-3)$ out of the set $\{\beta_0,\beta_1,\ldots,\beta_{D-2} \}$ are independent, giving $(D-1)$ free parameters to the solution. When $\eta=0$ and $z =1$, they reduce to the usual Kasner condition where both equations sum to unity. The modified Kasner condition with $\eta\neq0$ is similar to the modification in the space-like version considered in \cite{Awad:2007fj}.

Various special cases can be obtained by appropriate choices of the parameters. The following are a few examples: 
\begin{itemize}
 \item The AdS-Kasner solution of \cite{Ren:2016xhb} is recovered by setting $z=1$ and $\eta=0$. This switches off all the matter fields, leaving us with pure AdS gravity with the negative cosmological constant $\Lambda=-\half(D-1)(D-2)\ell^{-2}$.
 \item The Lifshitz black hole is obtained by setting $\beta_0=1$ and $\eta=0$. By the Kasner conditions \Eqref{KasnerConditions}, this forces all the $\beta_i$s to zero. 
 \item The planar AdS naked singularity \cite{Saenz:2012ga} is the case $z=1$, and the exponents at the spatial coordinates are equal such that
  \begin{align}
   \beta_0&=\frac{\nu(D-2)+1}{D-1},\quad\beta_1=\ldots=\beta_{D-2}=\frac{1-\nu}{D-1},
  \end{align}
  where the exponents are now parameterised by a single quantity $\nu\leq 1$. The scalar field is then reparametrised to
  \begin{align}
   \psi=\half\sqrt{\frac{(D-2)(1-\nu^2)}{D-1}}\ln f,
  \end{align}
  producing the solution the form given by the author in a previous paper \cite{Lim:2017dqw}.

\end{itemize}

To explore the IR geometry near $u\sim u_0$, we let $u=u_0-\bar{u}^2$. To leading order in $\bar{u}$, and upon rescaling $t$ and $x^i$ accordingly, the metric behaves like a radial, Ricci-flat Kasner geometry
\begin{align}
 \dif s^2_{\mathrm{IR}}\sim-\bar{u}^{2\beta_0}\dif t^2+\dif\bar{u}^2+\sum_i\bar{u}^{2\beta_i}\brac{\dif x^i}^2. \label{IR_geometry}
\end{align}
We see that the Kasner singularity will be absent if any one of the exponents are equal to $1$, making all others go to zero. In particular if $\beta_0=1$ and $\beta_1=\ldots=\beta_{D-2}=0$, this corresponds to the case of the Lifshitz black hole and Eq.~\Eqref{IR_geometry} for these parameters is simply its near-horizon geometry. If one of the spatial exponents $\beta_k=1$ for some $k$, this corresponds to the Lifshitz soliton which we will study in further detail in Sec.~\ref{LifSol}, and Eq.~\Eqref{IR_geometry} is then the metric expanded near the pole around the symmetry axis. On the other hand, the UV region is probed by $u\sim 0$. In this region, we have $f\sim1$ and the geometry is indeed that of Lifshitz,
\begin{align}
 \dif s^2_{\mathrm{UV}}\sim\ell^2\brac{-\frac{\dif t^2}{u^{2z}}+\frac{\dif u^2}{u^2}+\sum_i\frac{\brac{\dif x^i}^2}{u^2}}.
\end{align}

\section{The Null Energy Condition} \label{NEC}

The Null Energy Condition (NEC) requires that $R_{\mu\nu}k^\mu k^\nu\geq0$ for any null vector $k^\mu$. Here, we find it convenient to work in `domain-wall' coordinate $\rho=-\ln u$, in which the Ricci tensor components are calculated in Appendix \ref{Derivation}, where the function $f$ in this coordinate is
\begin{align}
 f=\expo{-2G},\quad G=-\half\ln\brac{1-c\expo{-(z+D-2)\rho}},\quad c=u_0^{-(z+D-2)}.
\end{align}
By using $k^\mu k_\mu=0$, we find that 
\begin{align}
 R_{\mu\nu}k^\mu k^\nu&=\sum_i\brac{\mathcal{E}_{tt}-\mathcal{E}_{ii}}\expo{2\rho-2\beta_i G}\brac{k^i}^2+\brac{\mathcal{E}_{tt}-\mathcal{E}_{\rho\rho}}\expo{2G}\brac{k^\rho}^2,
\end{align}
where $\mathcal{E}_{tt}$, $\mathcal{E}_{ii}$, and $\mathcal{E}_{\rho\rho}$ are as given in Eq.~\Eqref{Ecal_def} in Appendix \ref{Derivation}. Therefore, for the NEC to be satisfied, we require $\mathcal{E}_{tt}-\mathcal{E}_{ii}\geq 0$ and $\mathcal{E}_{tt}-\mathcal{E}_{\rho\rho}\geq 0$. We first look at the former, namely
\begin{align}
 \mathcal{E}_{tt}-\mathcal{E}_{ii}&=(z-1)(z+D-2)\sum_i\expo{2\rho-2\beta_i G}\geq 0, \label{NEC1}
\end{align}
leading to the usual constraint for the Lifshitz exponent where $z\geq 1$. The second inequality requires
\begin{align}
 \mathcal{E}_{tt}-\mathcal{E}_{\rho\rho}&=(z-1)(D-1)\expo{-2G}-2(z-1)(1-\beta_0)G'+KG'^2\geq0,\label{NEC2}
\end{align}
where
\begin{align}
 K=\frac{(z-1)(1-\beta_0)^2}{D-2}+\eta^2.
\end{align}
note that $-K$ correspond to the second and third terms in the right-hand side of Eq.~\Eqref{KasnerCondition2}, which appears upon invoking the Kasner conditions in the calculation of $R_{\rho\rho}$. This inequality will be satisfied if each of the terms in Eq.~\Eqref{NEC2} above are positive. The first term will be positive by $z\geq1$ as required by Eq.~\Eqref{NEC1}, while the second term  requires $\beta_0\leq 1$, since $G'$ is negative for $c\geq0$. 

The last term of Eq.~\Eqref{NEC2} is particularly interesting as it hints to us of the possibility of having a phantom scalar field that still satisfies the NEC. For this last term to be positive, we require $K\geq 0$, or
\begin{align}
 \eta^2\geq-\frac{(z-1)(1-\beta_0)^2}{D-2}, \label{eta_range}
\end{align}
which is immediately satisfied if $\eta$ is real. However, Eq.~\Eqref{eta_range} becomes non-trivial if $\eta$ is complex, or equivalently, $\psi$ is a phantom scalar. This is easy to see by an analytical continuation on $\psi$,
\begin{align}
 \psi\rightarrow\im\tilde{\psi}.
\end{align}
This gives us a phantom scalar, as the kinetic term of the action \Eqref{action} is changed to $\brac{\nabla\psi}^2\rightarrow-(\nabla\tilde{\psi})^2$. This only modifies the Einstein equation by $\nabla_\mu\psi\nabla_\nu\psi\rightarrow-\nabla_\mu\tilde{\psi}\nabla_\nu\tilde{\psi}$, with the Maxwell, dilaton, and scalar equations remain unchanged. Therefore, Eq.~\Eqref{LifKas_Solution} still solves the new Einstein equation if $\eta$ is replaced by
\begin{align}
 \eta\rightarrow\im\tilde{\eta},
\end{align}
for a new real parameter $\tilde{\eta}$. It follows that Eq.~\Eqref{eta_range} is modified to
\begin{align}
 \tilde{\eta}^2\leq \frac{(z-1)(1-\beta_0)^2}{D-2}. \label{phantom_eta_range}
\end{align}
Thus, a phantom scalar supporting the solution \Eqref{LifKas_Solution} can obey the NEC as long as its strength does not exceed the limit in \Eqref{phantom_eta_range}.

One of the main implications of this is that if $\tilde{\eta}$ saturates this condition, then Eqs.~\Eqref{KasnerConditions} reduces back to the original Kasner conditions where $K=0$ and
\begin{align}
 \beta_0+\sum_i\beta_i=\beta_0^2+\sum_i\beta_i^2=1.
\end{align}
This will be the key ingredient in constructing the Lifshitz soliton in a Sec.~\ref{LifSol} below.

\section{Geodesics} \label{Geodesics}

The geodesic structure of different special cases of the solution \Eqref{LifKas_Solution} have been studied in various works. For instance, geodesics of asymptotically Lifshitz spacetimes have been studied by \cite{Keeler:2013msa,Gentle:2015cfp}. The $D=3$ case in particular have been considered in \cite{Cruz:2013ufa,Villanueva:2013gra}. In the asymptotically AdS case, the effective potential of particles in the AdS-Kasner spacetime was analysed in Ref.~\cite{Ren:2016xhb}. The trajectories of particles in the AdS soliton were explicitly obtained in \cite{Shi:2016bxz}. In this section, we shall study the geodesic structure of the full solution given by our metric \Eqref{LifKas_Solution}.

A geodesic is described by a trajectory $x^\mu(\tau)$, where $\tau$ is an appropriate parametrisation along the curve. The Lagrangian for the geodesic motion is given by the invariant $L=\half g_{\mu\nu}\dot{x}^\mu\dot{x}^\nu\equiv\frac{\epsilon}{2}$, where over-dots denote derivatives with respect to $\tau$. The equations of motion is then determined from the Euler-Lagrange equations $\frac{\dif}{\dif\tau}\frac{\dif L}{\dif\dot{x}^\mu}=\frac{\dif L}{\dif x^\mu}$. For the metric described by \Eqref{LifKas_Solution}, the Lagrangian is
\begin{align}
 L&=\half\brac{-\frac{f^{\beta_0}\dot{t}^2}{u^{2z}}+\frac{\dot{u}^2}{u^2f}+\frac{1}{u^2}\sum_i f^{\beta_i}(\dot{x}^i)^2}=\frac{\epsilon}{2}. \label{GeodesicLagrangian}
\end{align}
By appropriately rescaling parameter $\tau$, we may fix the magnitude of $\epsilon$ to be unity if it is non-zero. Therefore we have the following cases 
\begin{align}
 \epsilon&=\left\{\begin{array}{rl}
                  -1, & \mbox{time-like geodesics,}\\
                  0, &\mbox{null/photon geodesics,}\\
                  +1, &\mbox{space-like geodesics.}
                  \end{array}\right.
\end{align}
In this paper, we will primarily focus on null and time-like geodesics. 

As $t$ and $x^i$ are cyclic variables of the Lagrangian, we have the first integrals
\begin{align}
 \dot{t}=\frac{Eu^{2z}}{f^{\beta_0}},\quad \dot{x}^i=\frac{P_iu^2}{f^{\beta_i}}, \label{ConstantsOfMotion}
\end{align}
where $E$ and $P_i$ are constants of motion which may be interpreted as the energy and momenta of the test particle.

Applying the Euler-Lagrange equation for $u$ gives us a second-order differential equation,
\begin{align}
 \ddot{u}&=\brac{\frac{1}{u}+\frac{f'}{2f}}\dot{u}^2-\brac{\frac{\beta_0 f'}{2f}-\frac{z}{u}}E^2+\sum_i\brac{\frac{\beta_if'}{2f}-\frac{1}{u}}P_i^2. \label{uddot}
\end{align}
However, by the invariance of $\epsilon=g_{\mu\nu}\dot{x}^\mu\dot{x}^\nu$, we can obtain a first-order equation for $u$, which is 
\begin{align}
 \dot{u}^2=u^{2z+2}f^{1-\beta_0}\brac{E^2-V^2_{\mathrm{eff}}}, \label{FirstIntegral}
\end{align}
where we have used \Eqref{ConstantsOfMotion} to eliminate $\dot{t}$ and $\dot{x}^i$ in favour of the constants, and $V^2_{\mathrm{eff}}$ is the effective potential
\begin{align}
 V_{\mathrm{eff}}^2=\frac{1}{u^{2z-2}}\sum_i f^{\beta_0-\beta_i}P_i^2-\frac{\epsilon f^{\beta_0}}{u^{2z}}.\label{V2_eff}
\end{align}

By inspection of the effective potential, null geodesics with $P_i\neq0$ and $z>1$ encounter an infinite potential barrier at $u\rightarrow0$. As pointed out in previous literature \cite{Keeler:2013msa,Gentle:2015cfp}, this is the consequence of the non-relativistic nature of the boundary, where the speed of light effectively becomes infinite. On the other hand, if $P_i\neq0$ for some $i$, the behaviour of the first term of \Eqref{V2_eff} depends on the exponent $\beta_0-\beta_i$. Since $f$ goes to zero for $u\rightarrow u_0$, the term either diverges for $\beta_i>\beta_0$, or vanishes for $\beta_i<\beta_0$. Therefore, in the former case, null geodesics will encounter another infinite barrier at $u\rightarrow u_0$, or in the latter case, the potential drops to zero. In the critical case $\beta_i=\beta_0$, the potential simply tends to $V^2_{\mathrm{eff}}\rightarrow P_i^2/u_0^{2z-2}$. 

An interesting situation arises if we combine the conditions for infinite barriers on both sides, namely
\begin{align}
 z>1,\quad \beta_i>\beta_0,\quad P_i\neq 0,\label{PhotonTrapCondition}
\end{align}
for some direction $i$. Then the geodesics will experience an infinitely deep potential well, implying that photons will be be trapped in a finite region of along the $u$-direction.

\begin{figure}
 \begin{center}
  \begin{subfigure}[b]{0.48\textwidth}
  \centering
  \includegraphics[scale=0.95]{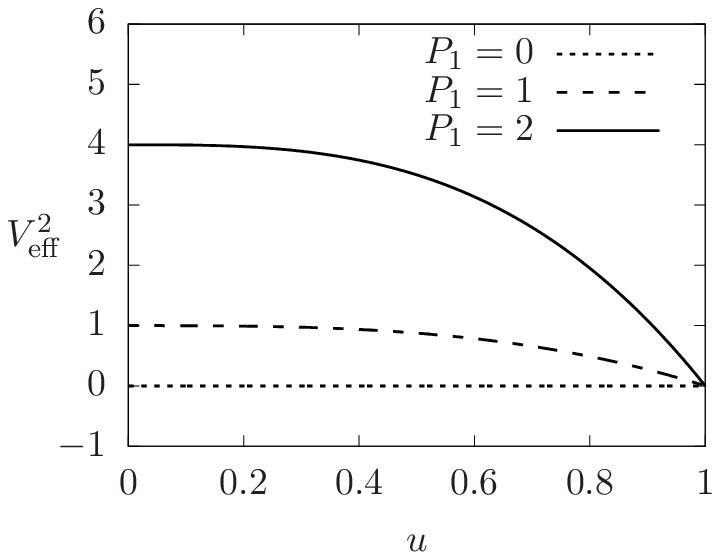}
  \caption{$\beta_0=1$, $\beta_1=0$, $z=1$.}
  \label{fig_data31x}
 \end{subfigure}
 \begin{subfigure}[b]{0.48\textwidth}
  \centering
  \includegraphics[scale=0.95]{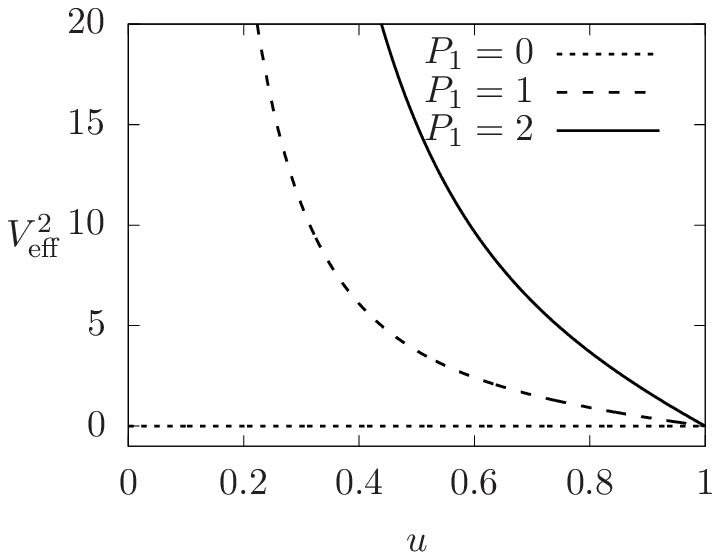}
  \caption{$\beta_0=1$, $\beta_1=0$, $z=2$.}
  \label{fig_data32x}
 \end{subfigure}
 \begin{subfigure}[b]{0.48\textwidth}
  \centering
  \includegraphics[scale=0.95]{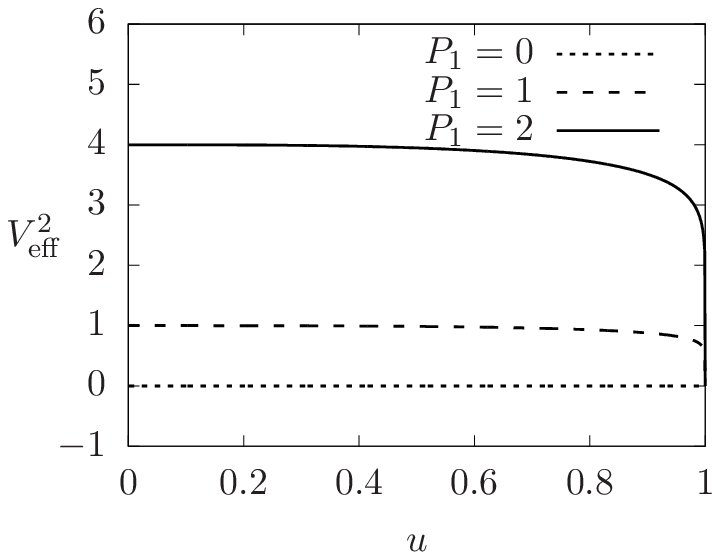}
  \caption{$\beta_0=0.6$, $\beta_1=0.5$, $z=1$.}
  \label{fig_data33x}
 \end{subfigure}
 \begin{subfigure}[b]{0.48\textwidth}
  \centering
  \includegraphics[scale=0.95]{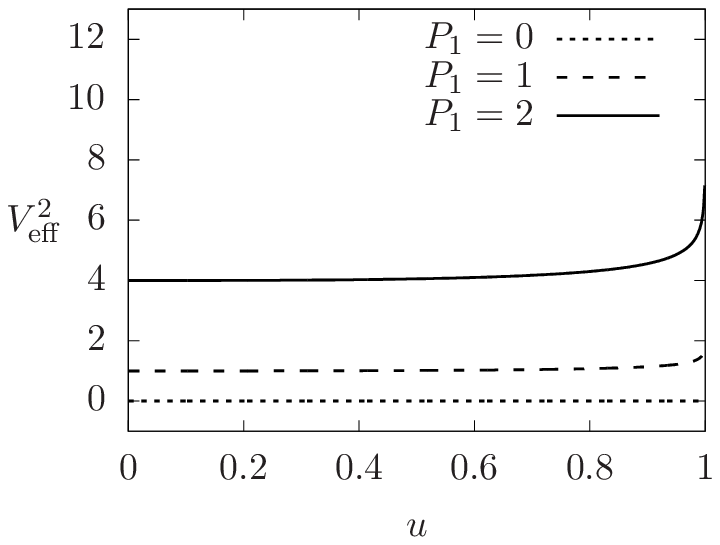}
  \caption{$\beta_0=0.2$, $\beta_1=0.8$, $z=1$.}
  \label{fig_data34x}
 \end{subfigure}
 \begin{subfigure}[b]{0.48\textwidth}
  \centering
  \includegraphics[scale=0.95]{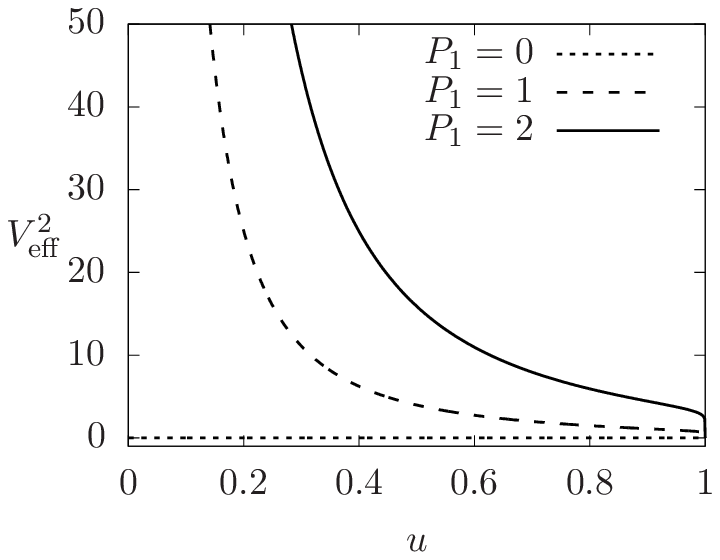}
  \caption{$\beta_0=0.6$, $\beta_1=0.5$, $z=2$.}
  \label{fig_data35x}
 \end{subfigure}
 \begin{subfigure}[b]{0.48\textwidth}
  \centering
  \includegraphics[scale=0.95]{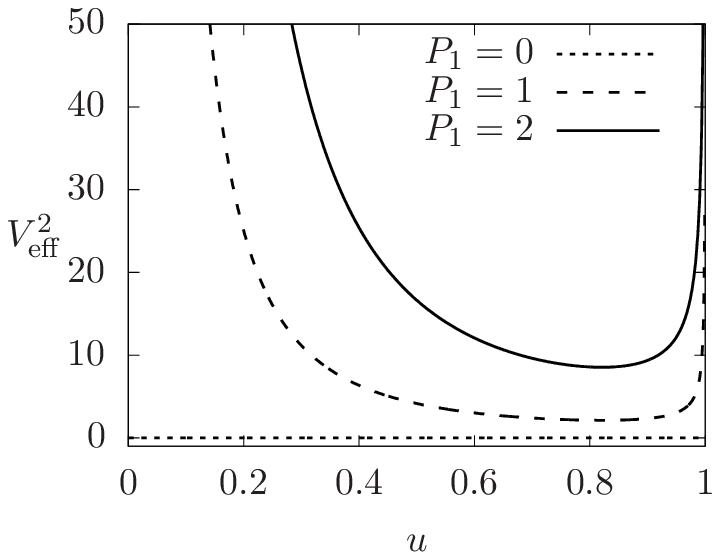}
  \caption{$\beta_0=0.2$, $\beta_1=0.8$, $z=2$.}
  \label{fig_data36x}
 \end{subfigure}
 \end{center}
 \caption{Plots of $V^2_{\mathrm{eff}}$ vs $u$ for $D=4$ and $u_0=1$, for null geodesics with $\epsilon=0$. From the Kasner condition, $\beta_2$ is determined by the first Kasner condition, $\beta_2=1-\beta_0-\beta_1$.}
 \label{fig_data3xx}

\end{figure}

In the case of $D=4$, our coordinates are $(t,x^1,x^2,u)$. The plots of the potentials for null geodesics in this case are shown in Fig.~\ref{fig_data3xx} with $u_0=1$, $\beta_2=0$, and $\beta_1=1-\beta_0$. In particular, we can see the $u=0$ barriers whenever $z>1$ (Figs.~\ref{fig_data32x}, \ref{fig_data35x}, and \ref{fig_data36x}), and the $u=u_0$ barriers for $\beta_1<\beta_0$ (Figs.~\ref{fig_data34x} and \ref{fig_data36x}). The $P_1=1$ and $P_1=2$ curves in Fig.~\ref{fig_data36x} are plotted with parameters that satisfy Eq.~\Eqref{PhotonTrapCondition}, therefore we have an infinite potential well with barriers on both $u=0$ and $u=u_0$. Photons with energy $E$ and momentum $P_i(\neq0$ for some $i$) are in a bound state oscillating within the range  $u_-\leq u\leq u_+$, where $u_\pm$ are roots of the equation $E^2=V^2_{\mathrm{eff}}$, corresponding to the turning points $\dot{u}=0$.

\begin{figure}
 \begin{center}
  \begin{subfigure}[b]{0.48\textwidth}
  \centering
  \includegraphics[scale=0.95]{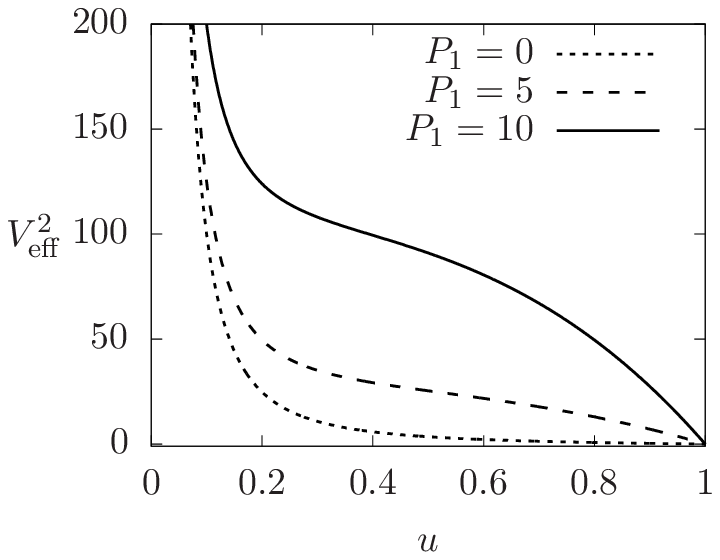}
  \caption{$\beta_0=1$, $\beta_1=0$, $z=1$.}
  \label{fig_data41x}
 \end{subfigure}
 \begin{subfigure}[b]{0.48\textwidth}
  \centering
  \includegraphics[scale=0.95]{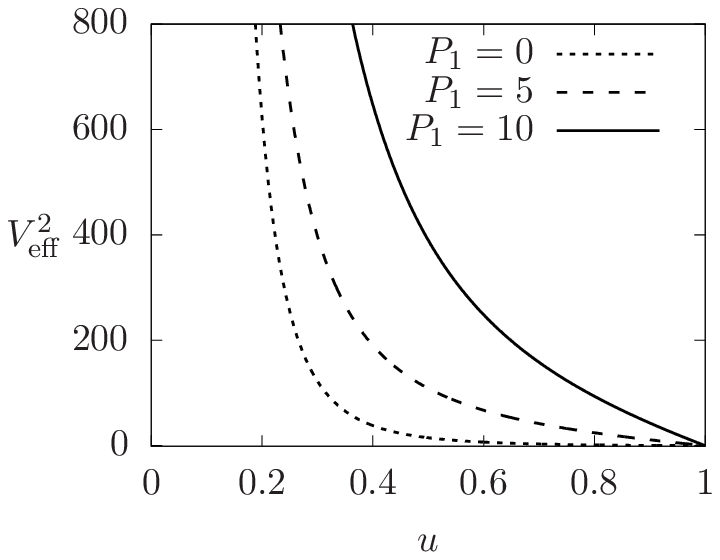}
  \caption{$\beta_0=1$, $\beta_1=0$, $z=2$.}
  \label{fig_data42x}
 \end{subfigure}
 \begin{subfigure}[b]{0.48\textwidth}
  \centering
  \includegraphics[scale=0.95]{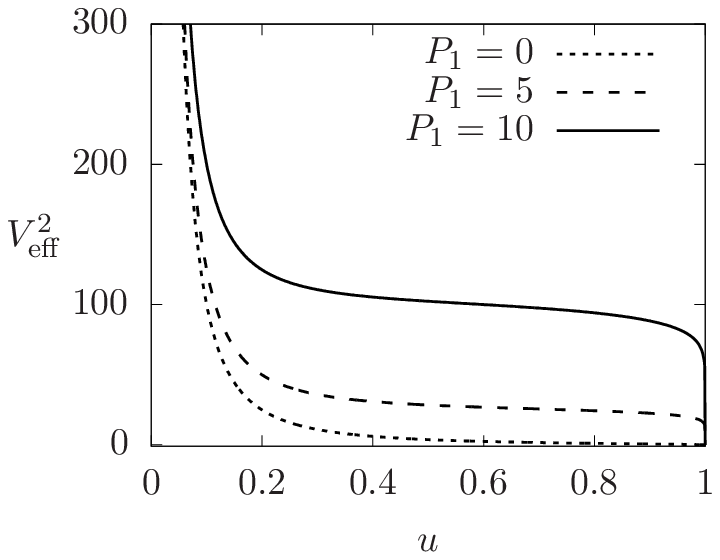}
  \caption{$\beta_0=0.6$, $\beta_1=0.5$, $z=1$.}
  \label{fig_data43x}
 \end{subfigure}
 \begin{subfigure}[b]{0.48\textwidth}
  \centering
  \includegraphics[scale=0.95]{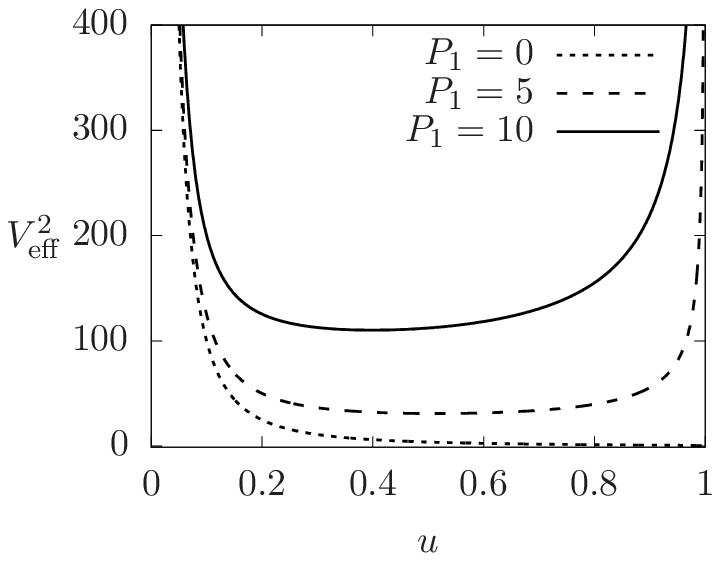}
  \caption{$\beta_0=0.2$, $\beta_1=0.8$, $z=1$.}
  \label{fig_data44x}
 \end{subfigure}
 \begin{subfigure}[b]{0.48\textwidth}
  \centering
  \includegraphics[scale=0.95]{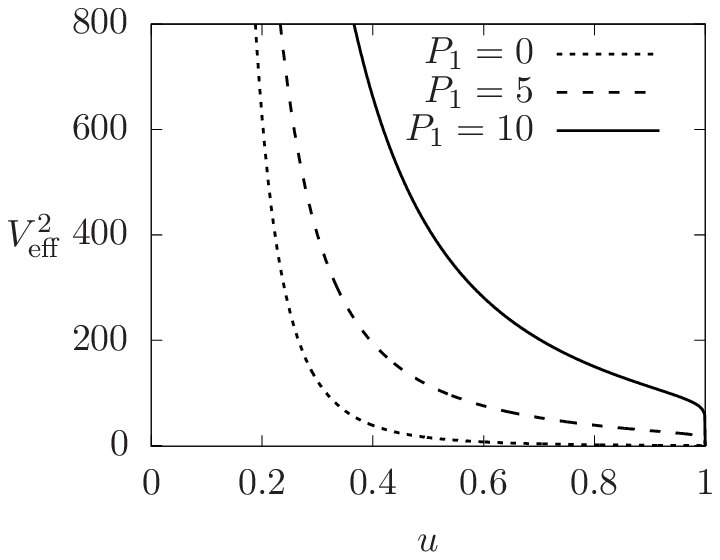}
  \caption{$\beta_0=0.6$, $\beta_1=0.5$ $z=2$.}
  \label{fig_data45x}
 \end{subfigure}
 \begin{subfigure}[b]{0.48\textwidth}
  \centering
  \includegraphics[scale=0.95]{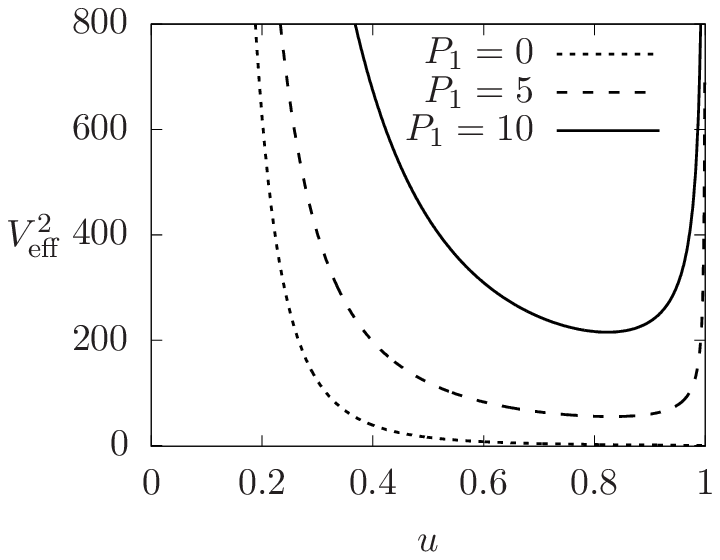}
  \caption{$\beta_0=0.2$, $\beta_1=0.8$ $z=2$.}
  \label{fig_data46x}
 \end{subfigure}
 \end{center}
 \caption{Plots of $V^2_{\mathrm{eff}}$ vs $u$ for $D=4$ and $u_0=1$ for time-like geodesics with $\epsilon=-1$. From the Kasner condition, $\beta_2$ is determined by the first Kasner condition, $\beta_2=1-\beta_0-\beta_1$.}
 \label{fig_data4xx}

\end{figure}
For time-like geodesics with $\epsilon=-1$, the non-zero second term in \Eqref{V2_eff} implies that there will always be an infinite barrier at $u\rightarrow 0$ for any $z\geq 1$, hence this includes the well-known AdS case. Similar to the null case, if the particle has a non-zero transverse momentum in spacetimes with $\beta_i>\beta_0$ for some $i$, there will be an infinite barrier at $u\rightarrow u_0$ as well. Hence, the condition
\begin{align}
 \beta_i>\beta_0,\quad P_i\neq 0 \label{TimelikeTrapCondition}
\end{align}
for some $i$ will also lead to time-like particles being trapped in an infinite potential well. Similar to the description for the trapped photons, time-like particles with energy $E$ and some non-zero $P_i$ will also oscillate between the turning points $E^2=V^2_{\mathrm{eff}}$. In the AdS case where $z=1$, the condition \Eqref{TimelikeTrapCondition} contains the AdS soliton case $\beta_1=1$, thus includes the results of \cite{Shi:2016bxz} as a special case. 

The potentials for time-like particles in $D=4$ are shown in Fig.~\ref{fig_data4xx} for $u_0=1$, $\beta_2=0$, and $\beta_1=1-\beta_0$. Here we can see that there is always a potential barrier at $u=0$, while another barrier develops at $u=u_0$ for $\beta_1>\beta_0$. As depicted by examples in Fig.~\ref{fig_data44x} and \ref{fig_data46x}, this corresponds to a potential well with infinite barriers on both sides.

\section{Lifshitz soliton} \label{LifSol}

The authors of \cite{Lu:2013tza} have studied holographic superconductivity in the background of a Lifshitz soliton, which they have constructed by performing a double-Wick rotation on a Lifshitz black hole. Thus, this can be considered an analogue to the AdS soliton where the metric, in the notation of the present paper, is \cite{Lu:2013tza}
\begin{align}
 \dif s^2&=\ell^2\sbrac{\frac{1}{u^2}\brac{-\dif\tau^2+\frac{\dif u^2}{f}+\dif\vec{x}^2_{(D-3)}}+\frac{f\dif\sigma^2}{u^{2z}}},\nonumber\\
      f&=1-\brac{\frac{u}{u_0}}^{D-1},
\end{align}
where we have denoted $\dif\vec{x}^2_{(D-3)}=\sum_{i=2}^{D-2}\brac{\dif x^i}^2$. By the Wick rotation, we see that the coordinate that carries the Lifshitz scaling has turned into a spatial one. The cost of this procedure is that the $u\rightarrow0$ boundary of this metric is no longer has the non-relativistic property in the sense that the Lorentz invariance between space and time being broken. 

\subsection{The solution and removal of its conical singularity}

In this section, we will obtain another analogue to the AdS soliton which still preserves the Lifshitz scaling in the time coordinate. Instead of performing a double-Wick rotation, the solution is extracted as a special case of \Eqref{LifKas_Solution} where $\psi$ is turned into a phantom scalar by $\eta\rightarrow\im\tilde{\eta}$.

An obvious way to obtain the Lifshitz soliton from \Eqref{LifKas_Solution} is by setting $\beta_0$, and all but one of the $\beta_i$'s to zero. Let us single out $\beta_1$ to remain non-zero, and have
\begin{align}
 \beta_0=\beta_2=\ldots=\beta_{D-2}=0,\quad\beta_1=1,\quad x^1\equiv\sigma.
\end{align}
The first Kasner condition is trivially satisfied, while the second one is, with $\eta\rightarrow\im\tilde{\eta}$,
\begin{align}
 \tilde{\eta}^2=\frac{z-1}{D-2}.
\end{align}
This value of $\tilde{\eta}$ just saturates Eq.~\Eqref{phantom_eta_range} with $\beta_0=0$, thus the NEC is not violated. Writing out the metric for this solution, we have
\begin{align}
 \dif s^2&=\ell^2\sbrac{-\frac{\dif t^2}{u^{2z}}+\frac{1}{u^2}\brac{\dif\vec{x}^2_{(D-3)}+\frac{\dif u^2}{f}+f\dif\sigma^2}},\nonumber\\
     f&=1-\brac{\frac{u}{u_0}}^{z+D-2}.
\end{align}
Just like the soliton limit of the AdS-Kasner solution, the metric at $u=u_0$ is no longer a curvature singularity, nor it is an event horizon. It is simply a conical singularity which can be removed by imposing a periodicity on coordinate $\sigma$ by
\begin{align}
 \sigma\sim\sigma+\frac{2\pi}{\kappa}, \label{ImposePeriodicity}
\end{align}
where $\kappa$ is given by
\begin{align}
 \kappa=\half\left|\frac{\dif f}{\dif u} \right|_{u=u_0}=\frac{z+D-2}{2u_0}. \label{EuclideanSurfaceGravity}
\end{align}
Therefore, the location $u=u_0$ may be interpreted as a pole representing the tip of a cigar-shaped geometry.

\subsection{Bound photon orbits}

Since the parameters of the Lifshitz soliton satisfy the condition \Eqref{PhotonTrapCondition}, photons with non-zero transverse momentum in this spacetime will be bound within an infinitely deep potential well. Bound photon orbits within a potential well are comparatively novel, as it is more common for most (especially asymptotically flat) spacetimes to only have unstable circular photon orbits. In this subsection, we shall explore the nature of these trajectories in further detail. For the specific parameters of the Lifshitz soliton ($\beta_0=\beta_2=\ldots=\beta_{D-2}=0$), we reproduce the geodesic equations here for convenient reference:
\begin{align}
 \dot{t}&=Eu^{2z},\quad \dot{\sigma}=\frac{Pu^2}{f},\quad \dot{x}^j=p_ju^2,\quad j=2,\ldots, D-2,\nonumber\\
 \dot{u}^2&=u^{2z+2}f\brac{E^2-V^2_{\mathrm{eff}}},\quad V^2_{\mathrm{eff}}=\frac{P^2}{u^{2z+2}f}+\frac{1}{u^{2z+2}}\sum_{j}p_j^2-\frac{\epsilon}{u^{2z}}, \label{LifSol_Geodesics}
\end{align}
where we have let $P_1=P$, and $P_j=p_j$ for $j=2,\ldots,D-2$ to distinguish the momenta in the compactified $x^1=\sigma$ direction. As mentioned in the previous subsection, the location $u=u_0$ is simply interpreted as the pole at the tip of a cigar-shaped geometry. In light of this, the infinite potential barrier at $u=u_0$ can now be interpreted as a behaviour analogous to ordinary Keplerian motion, where particles with non-zero angular momentum are not able to reach the symmetry axis of the gravitating source.

\begin{figure}
 \begin{center}
  \includegraphics{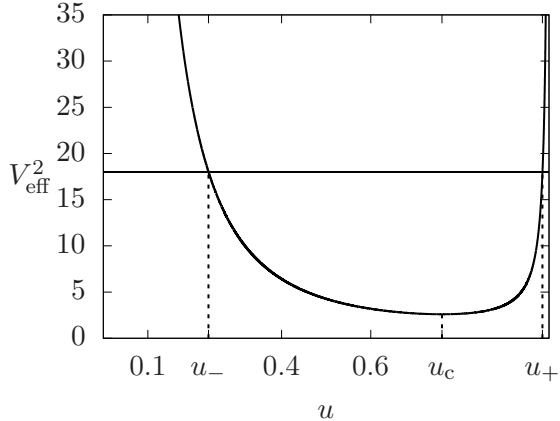}
 \end{center}
 \caption{Effective potential of null ($\epsilon=0$) geodesics in the Lifshitz soliton spacetime with $D=4$, $u_0=1$. The momentum of the trajectory is $P_1\equiv P=1$, and $P_2\equiv p_2=0$. The horizontal line indicates the particle energy $E^2=18$. For these parameters, $u_-=0.23607$, $u_+=0.98538$, and $u_{\mathrm{c}}=0.75984$.}
 \label{fig_LifSolEffPot}
\end{figure}

As we are particularly interested in bound null geodesics, we shall set $\epsilon=0$, and, for simplicity, also set $p_j=0$. If the photon has $P\neq0$, then it satisfies the condition \Eqref{PhotonTrapCondition}. As seen in Fig.~\ref{fig_LifSolEffPot}, the shape of the potential well is typically asymmetric, where the $u=u_0$ barrier rises more sharply than the $u=0$ side. Photon of fixed energy $E$ will be bound within the range $u_-\leq u\leq u_+$ where $u_\pm$ are the turning points in the $u$-motion. From \Eqref{LifSol_Geodesics}, they can be obtained as roots of the equation $E^2=V^2_{\mathrm{eff}}$. This photon is able to come very close the pole where it attains very high `angular velocity' in the sense that $|\dot{\sigma}|=|Pu^2/f|$ is large due to the fact that $f$ is small near the pole. At the same time, $\dot{u}$ very quickly transitions from its highest value near the potential minimum to its turning point $\dot{u}=0$ where $E^2=V^2_{\mathrm{eff}}$. So it is quickly bounced back outwards by the barrier. The potential has a minimum at
\begin{align}
 u=u_{\mathrm{c}}=\exp\sbrac{\frac{(z+D-2)\ln u_0+\ln\brac{\frac{2(z-1)}{3z+D-4}}}{z+D-2}},
\end{align}
for which $\frac{\dif\brac{V^2_{\mathrm{eff}}}}{\dif u}=0$ at $u=u_{\mathrm{c}}$. Therefore, photons with energy $E^2=V^2_{\mathrm{eff}}(u_\mathrm{c})$ are in a stable `circular' orbit at constant $u$. The effective potential a photon with $P=1$ is plotted in Fig.~\ref{fig_LifSolEffPot}, where a particle of energy $E^2=18$ can be seen to be bound between $u_-\leq u\leq u_+$, along with the location of the potential minimum $u_{\mathrm{c}}$.

\begin{figure}
 \begin{center}
  \begin{subfigure}[b]{0.48\textwidth}
    \centering
    \includegraphics[scale=1]{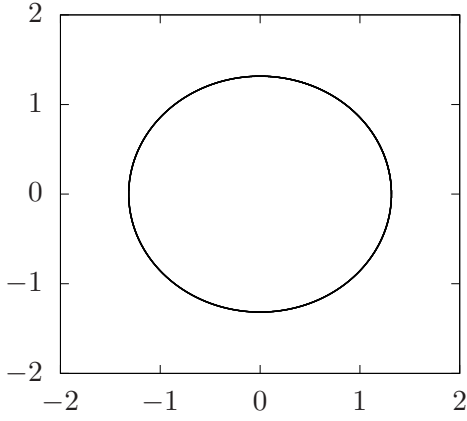}
    \caption{`Circular' orbit.}
    \label{fig_circular}
  \end{subfigure}
  \begin{subfigure}[b]{0.48\textwidth}
    \centering
    \includegraphics[scale=1]{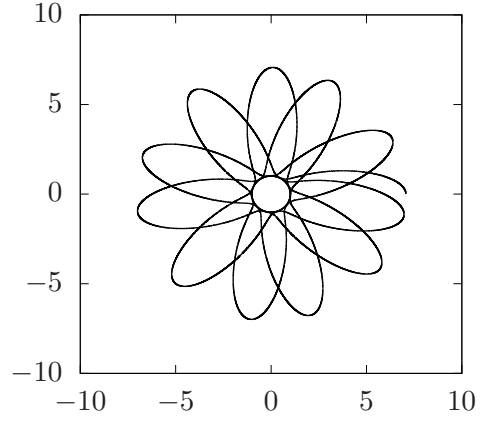}
    \caption{Generic orbit.}
    \label{fig_generic}
  \end{subfigure}
  \begin{subfigure}[b]{0.48\textwidth}
    \centering
    \includegraphics[scale=1]{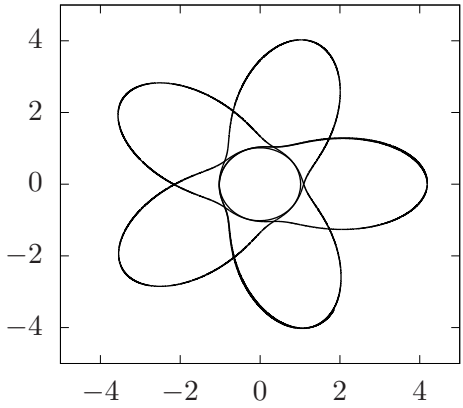}
    \caption{Periodic, $\mathcal{V}=4$, $\mathcal{Z}=5$, $q=\frac{\mathcal{V}}{\mathcal{Z}}=\frac{4}{5}$.}
    \label{fig_periodic_n5}
 \end{subfigure}
 \begin{subfigure}[b]{0.48\textwidth}
    \centering
    \includegraphics[scale=1]{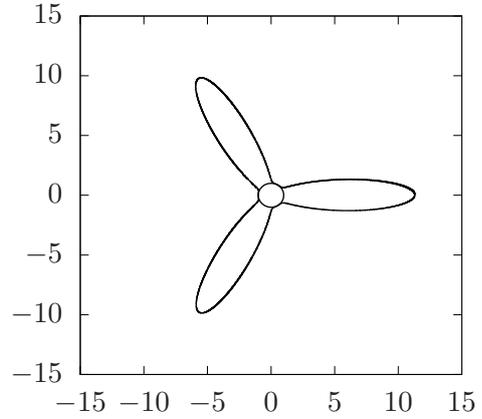}
    \caption{Periodic, $\mathcal{V}=2$, $\mathcal{Z}=3$, $q=\frac{\mathcal{V}}{\mathcal{Z}}=\frac{2}{3}$.}
    \label{fig_periodic_n6}
 \end{subfigure}
 \end{center}
 \caption{Plots of various photon trajectories in Cartesian-like coordinates \Eqref{CartesianLikeCoords} in the $D=4$ Lifshitz soliton spacetime with $u_0=1$. All the trajectories have momentum $P=1$, while the energies are $E^2=2.59808$ for the circular orbit (Fig.~\ref{fig_circular}), $E^2=50$ for the generic orbit (Fig.~\ref{fig_generic}), $E^2=17.601$ for the $q=\frac{4}{5}$ orbit (Fig.~\ref{fig_periodic_n5}), and $E^2=127.98$ for the $q=\frac{2}{3}$ orbit (Fig.~\ref{fig_periodic_n6}).}
 \label{fig_bound}
\end{figure}

The trajectories can be obtained by numerical integration of \Eqref{LifSol_Geodesics} to obtain the trajectory $u(\tau)$ and $\sigma(\tau)$.\footnote{In practice, we obtained $u(\tau)$ by integrating the equivalent equation \Eqref{uddot} using the fourth-order Runge-Kutta method.} An intuitive visualisation of the orbits can be presented using Cartesian-like coordinates,
\begin{align}
 X=\frac{1}{u}\cos\brac{\kappa\sigma},\quad Y=\frac{1}{u}\sin\brac{\kappa\sigma}, \label{CartesianLikeCoords}
\end{align}
where $\kappa$ given by Eq.~\Eqref{EuclideanSurfaceGravity}. Some examples are plotted in these coordinates in Fig.~\ref{fig_bound}.

We find that these trapped photons can be classified using a scheme developed by Levin and Perez-Giz \cite{Levin:2008mq}. This scheme is based on Poincar\'{e}'s paradigm where an understanding of a dynamical system can structured around its periodic orbits. Following the methods of \cite{Levin:2008mq}, a periodic null orbit can be found in our present system by considering the ratio of the frequencies in the $u$ and $\sigma$ directions. This can be obtained by integrating the quantity $\frac{\dif u}{\dif\sigma}$. Using Eq.~\Eqref{LifSol_Geodesics}, the integration results in
\begin{align}
 \Delta\sigma=\int_{0}^{\Delta\sigma}\dif\sigma=2\int_{u_-}^{u_+}\frac{P\;\dif u}{u^{z-1}f^{3/2}\sqrt{E^2-V^2_{\mathrm{eff}}}}.
\end{align}
The periodicity can be quantified by a number $q$ defined by
\begin{align}
 q=\frac{\kappa}{2\pi}\Delta\sigma.
\end{align}

Periodic orbits correspond to those where $q$ is a rational number. Furthermore, for these periodic orbits, we can write $q=\frac{\mathcal{V}}{\mathcal{Z}}$ for two integers $\mathcal{V}$ and $\mathcal{Z}$. To briefly review the scheme of \cite{Levin:2008mq}, $\mathcal{Z}$ is the number of leaves traced out by the periodic orbit, and $\mathcal{V}$ is the order in which the leaves are traced while completing the cycle. Unlike the Schwarzschild and Kerr orbits, the photon trajectories in the Lifshitz soliton spacetime does not `whirl' near the $u_+$ limit before going out towards $u_-$ again. So the whirl number for these photons are zero. Examples of periodic orbits for $q=\frac{4}{5}$ and $q=\frac{2}{3}$ are shown in Figs.~\ref{fig_periodic_n5}, and \ref{fig_periodic_n6}, respectively.

\section{Parameter space for \texorpdfstring{$D=4$}{D=4}} \label{Parameters}

Let us now consider the parameter ranges of the full solution in further detail. As we have seen in the previous sections, the parameters of the solution are constrained by the field equations and the NEC, which result in Eqs.~\Eqref{KasnerConditions} and \Eqref{eta_range}, or \Eqref{phantom_eta_range} if the scalar field is phantom. We will also find it convenient to combine Eqs.~\Eqref{eta_range} and \Eqref{phantom_eta_range} into a single inequality 
\begin{align}
 \xi\geq\xi_*,\quad \xi_*\equiv-\frac{(z-1)(1-\beta_0)^2}{D-2}, \label{xi_range}
\end{align}
where 
\begin{align}
 \xi=\left\{\begin{array}{cc}
            \eta^2, & \mbox{if }\xi\geq0,\\
            \brac{\im\tilde{\eta}}^2=-\tilde{\eta}^2, & \mbox{if }\xi<0.
            \end{array}\right.
\end{align}
Therefore we can let $\xi$ run from positive to negative and its sign indicating whether the scalar is non-phantom or otherwise.

A deeper grasp of the ranges can be obtained by considering the concrete case of $D=4$. In this case we have three exponents $(\beta_0,\beta_1,\beta_2)$. We can let $\beta_0$ and $\beta_1$ be free parameters and have the first Kasner condition \Eqref{KasnerCondition1} fix $\beta_2=1-\beta_0-\beta_1$. Combining the second Kasner condition \Eqref{KasnerCondition1} and our combined NEC inequality \Eqref{xi_range} gives
\begin{align}
 \xi-\xi_*=1-\sbrac{\beta_0^2+\beta_1^2+(1-\beta_0-\beta_1)^2}\geq0. \label{xi_range1}
\end{align}
The range of $(\beta_0,\beta_1)$ that satisfies this inequality is
\begin{subequations}\label{beta_range}
\begin{align}
 \frac{1-2\beta_0-\sqrt{1+4\beta_0-4\beta_0^2}}{2}&\leq\beta_1\leq\frac{1-2\beta_0+\sqrt{1+4\beta_0-4\beta_0^2}}{2},\\
 \frac{1-\sqrt{2}}{2}&\leq\beta_0\leq\frac{1+\sqrt{2}}{2}.
\end{align}
\end{subequations}
\begin{figure}
 \begin{center}
  \includegraphics{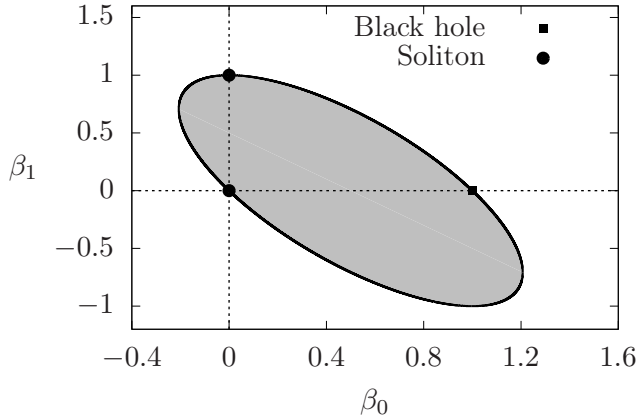}
 \end{center}
 \caption{Parameter ranges of the Kasner exponents $(\beta_0,\beta_1)$ in $D=4$. The shaded regions correspond to allowed values that satisfy the NEC, and the boundary curve is where it NEC inequality is saturated. The square point marks the value $\beta_0=1$, corresponding to the (Lifshitz or AdS) black hole, and the circular points mark the values for the (Lifshitz/AdS) soliton. }
 \label{fig_range_beta}

\end{figure}

The allowed ranges of $(\beta_0,\beta_1)$ is visualised in Fig.~\ref{fig_range_beta}. In the figure, the shaded regions correspond to the values of $(\beta_0,\beta_1)$ satisfying Eq.~\Eqref{beta_range}. Therefore the boundary of this shaded region is where $\xi=\xi_*$. There are three special points marking the values corresponding to metrics without naked singularities. Namely, the square point $(1,0)$ corresponds to the Lifshitz black hole. (Or, if $z=1$, the AdS black hole.) This point lies on the boundary curve because $\xi$ and $\xi_*$ are separately zero, corresponding to the trivial saturation of \Eqref{xi_range1}. The two circular points correspond to the regular Lifshitz (or AdS) solitons. The point $(0,1)$ requires $x^1=\sigma$ to be imposed with periodicity as shown in Eq.~\Eqref{ImposePeriodicity}. Thus $x^1$ becomes orbits around the axis of the `cigar' geometry. On the other hand, the point $(0,0)$ implies $\beta_2=1$ instead. Therefore it is $x^2$ that must be imposed with the periodicity instead. Unless $z=1$, these two soliton points require a non-zero $\xi_*$, and hence it non-trivially saturates $\xi\leq\xi_*$.

\begin{figure}
 \begin{center}
  \includegraphics{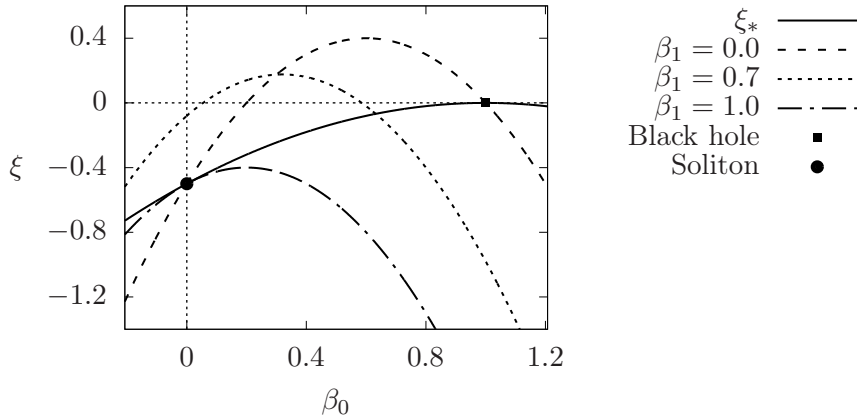}
 \end{center}
 \caption{Plot of $\xi$ as a function of $\beta_0$ for $z=2$ and various choices of $\beta_1$. The solid curve corresponds to the saturation $\xi=\xi_*$ and represents the lowest value of $\xi$ that does not violate the NEC.}
 \label{fig_range_xi}
\end{figure}

We can also see which parameter values require a phantom scalar by checking if $\xi$ is negative in Eq.~\Eqref{xi_range1}. In Fig.~\ref{fig_range_xi}, we show an example for $z=2$ as plots of $\xi$ against $\beta_0$ for various values of $\beta_1$. The solid line in Fig.~\ref{fig_range_xi} correspond to the saturation $\xi=\xi_*$. Therefore, for a given choice of $\beta_1$ and $z$, the function $\xi(\beta_0)$ must lie above this curve to satisfy the NEC. As in the previous figure, the square and circular points mark the parameters for the black hole and soliton, respectively. In particular, the $\beta_1=0$ curve intersects the $\xi=\xi_*$ curve at a the black-hole and soliton points, as the former has $\beta_0=1$, and the latter is at $\beta_0=0$, thus implying the $\beta_2=1$ soliton represented by $(0,0)$ in Fig.~\ref{fig_range_beta}. Thus, for $\beta_1$ fixed at zero, varying $\beta_0$ interpolates between the black hole and the soliton solution without violating the NEC. (Recall that $\beta_2$ accordingly varies through the Kasner condition.) Furthermore, we see that for $\beta_1=0$, the values of $\frac{1}{5}\leq \beta_0\leq 1$ require $\xi\geq0$, thus requiring a non-phantom scalar, while for $0\leq\beta_0<\frac{1}{5}$ a phantom scalar with $\xi<0$ is required to support this metric. On the other hand, the $\beta_1=1$ curve is always less than $\xi_*$, except where it is equal to $\xi_*$ at the soliton corresponding to the point $(0,1)$ in Fig.~\ref{fig_range_beta}. 

\section{Conclusion}\label{Conclusion}

We have presented an exact solution in Einstein-Maxwell-dilaton gravity corresponding to a spacetime with a Kasner-type singularity with Lifshitz asymptotics. This solution easily accommodates the addition of a new massless scalar field. We can regard this as a generalisation of the AdS-Kasner solution \cite{Ren:2016xhb} to include Lifshitz asymptotics and contains various known solutions as special cases.

This additional scalar field serves to modify the second Kasner condition further. Analytical continuation of this scalar essentially turns it onto a phantom scalar with an opposite sign of the kinetic term. We have found that the solution with a phantom scalar still satisfies the Null Energy Condition up to a certain limit of the field strength. Even though the NEC is satisfied, there is the potentially problematic issue in the quantum interpretation of the phantom scalar; namely that the wrong-sign kinetic term implies that the energy of the scalar is unbounded from below.

We have also studied the geodesic structure of this solution, where much of its qualitative aspects can be gleaned from the effective potential. Time-like particles encounter an infinite barrier at the boundary for any $z$, including the AdS case $z=1$. This is a well-known property where AdS acts as a confining potential for massive particles. For $z>1$, null geodesics with non-zero transverse momentum also encounter an infinite barrier at the boundary. This reflects the non-relativistic nature of the boundary where the speed of light is infinite. For spacetimes with $\beta_i>\beta_0$ for some direction $i$, there is another infinite barrier at $u=u_0$ for time-like and null particles. This is reminiscent of geodesics in the Fisher/Janis-Newman-Winicour spacetime where an infinite barrier also appears to particles with non-zero angular momentum if the scalar field supporting the solution is sufficiently strong \cite{Zhou:2014jja,Joshi:2013dva,Babar:2015kaa,Babar:2017gsg}.

We have focused on geodesics in the Lifshitz soliton spacetime, partly because this is an interesting case which is free from curvature singularities, and thus seems more physically relevant. Since it retains the asymptotically Lifshitz property for $u\rightarrow 0$, there is an infinite potential barrier for photons there. The location $u\rightarrow u_0$ in the Lifshitz soliton is no longer a curvature singularity, but rather a conical singularity which can be removed by imposing a periodicity on the coordinate $x^1=\sigma$. Just as in the AdS soliton solution, the location $u=u_0$ is now interpreted as the pole representing the tip of a cigar-shaped geometry. The interpretation potential barrier at $u=u_0$ also becomes straightforward: It is the analogue to a similar situation in ordinary Keplerian motion, where particles with non-zero angular momentum are unable to reach the symmetry axis of the gravitating source. Having an effective potential with infinite barriers at both sides, we explore the orbits of photon bound within this well. While a full taxonomy classification of the orbits is not attempted in this paper, we have demonstrated that periodic orbits can analysed under the procedure developed by Levin and Perez-Giz \cite{Levin:2008mq} for time-like particles around Kerr/Schwarzschild black holes. One notable difference in our case is that the photons do not tend to whirl near the `centre' of motion as relativistic time-like particles are wont to do in Kerr/Schwarzschild geometry.

In this paper, we have treated the solution purely within the context of $D=(d+1)$-dimensional Einstein-Maxwell-dilaton classical gravity with an additional massless, minimally-coupled scalar. As spacetimes with Lifshitz asymptotics might be a useful tool to condensed matter physics via the holographic correspondence, the $d$-dimensional boundary dual quantities of this spacetime might be of interest. Some of these quantities might further constrain the parameters of the solution. For instance, in the AdS-Kasner case, energy conditions for the CFT dual stress tensor imposes an additional constraint $\beta_0>\beta_i$ \cite{Ren:2016xhb}. It might be worth checking similar quantities for the solution \Eqref{LifKas_Solution} with Lifshitz asymptotics, especially in light of subtleties involving the boundary for these spacetimes \cite{Gentle:2015cfp}.

\section*{Acknowledgements}
The author thanks Dr. Qing-hai Wang for helpful comments and discussions.

\appendix 
\section{Derivation of the solution} \label{Derivation}

To derive the main solution presented in Sec.~\ref{Solution}, we begin with a metric ansatz which contains $(D-1)$ orthogonal Killing vectors. This is a special case of the class of generalised Weyl metrics \cite{Emparan:2001wk} and is written as
\begin{align}
 \dif s^2&=\ell^2\brac{\sum_{a=0}^{D-2}\epsilon_a\expo{2U_a}\brac{\dif x^a}^2+\expo{2G}\dif\rho^2},
\end{align}
where we assume the functions $U_0,\ldots,U_{D-2}$, and $G$ depend only on coordinate $\rho$. Here, $\epsilon_a=\pm 1$ depending on whether the Killing vector $\partial_a$ is time-like or space-like. Therefore, our natural convention for Lorentzian spacetimes would be $\epsilon_0=-1$ and $\epsilon_i=+1$ for $i=1,\ldots, D-2$. The Ricci tensor components for this metric is given by 
\begin{align}
 R_{aa}&=-\epsilon_a\expo{2U_a-2G}\sbrac{U_a''+U_a'\brac{\sum_{c=0}^{D-2} U_c'-G'}},\quad a=0,\ldots,D-2,\\
 R_{\rho\rho}&=-\sum_{a=0}^{D-2}\brac{U_a''+U_a'^2-G'U'_a},
\end{align}
where primes denote derivatives with respect to $\rho$.

In the following, we shall let $x^0=t$, along with $\epsilon_0=-1$. Our aim is to generalise the asymptotically AdS Kasner singularity to include Lifshitz asymptotics. Therefore, a natural choice which includes both metrics as special cases would be 
\begin{align}
 G&=-\half\ln\brac{1-c\expo{-(z +D-2)\rho}},\quad U_0=z \rho-\beta_0G,\quad U_i=\rho-\beta_iG,\quad i=1,\ldots,D-2. \label{LifKas_Functions}
\end{align}
We now seek the matter fields that supports a metric with this choice \Eqref{LifKas_Functions}. We begin with the following ansatz for the gauge field
\begin{align}
 A=\chi\,\dif t,
\end{align}
where $\chi$ is a function that depends only on $\rho$. If the scalar fields $\varphi$ and $\psi$ take the form (also assuming they depend only on $\rho$)
\begin{align}
 \alpha\varphi=(D-2)\rho-(1-\beta_0)G, \quad \psi=\eta G, \label{LifKas_Scalars}
\end{align}
for a constant $\eta$, the Maxwell and dilaton equations are solved to give 
\begin{align}
 \chi&=\frac{\ell e}{z +D-2}\expo{(z +D-2)\rho},\quad (z +D-2)(D-2)=2\alpha^2e^2,
\end{align}
where $e$ is an integration constant. (We have neglected the second integration constant which is a trivial additive to $\chi$.) With these results for $\varphi$ and $\chi$, the Einstein equations reduce to
\begin{align}
 -\mathcal{E}_{tt}&=\frac{2\Lambda\ell^2-2(D-3)e^2}{D-2},\label{LifKas_EE1}\\
 -\mathcal{E}_{ii}&=\frac{2\Lambda\ell^2-2(D-3)e^2}{D-2},\label{LifKas_EE2}\quad i=1,\ldots, D-2,\\
 -\mathcal{E}_{\rho\rho}&=\frac{2\Lambda\ell^2-2(D-3)e^2}{D-2}+\expo{-2G}\brac{\varphi'^2+\psi'^2}, \label{LifKas_EE3}
\end{align}
where we have abbreviated
\begin{subequations}\label{Ecal_def}
\begin{align}
 \mathcal{E}_{tt}&=\expo{-2G}\sbrac{\beta_0G''+(z -\beta_0G')\brac{z +D-2-\brac{1+\beta_0+\sum_i\beta_i}G'}},\\
 \mathcal{E}_{ii}&=\expo{-2G}\sbrac{\beta_iG''+(1-\beta_iG')\brac{z +D-2-\brac{1+\beta_0+\sum_i\beta_i}G'}},\\
 \mathcal{E}_{\rho\rho}&=\sbrac{-\beta_0G''+(z -\beta_0G')^2-G'(z -\beta_0G')}\nonumber\\
       &\quad+\sum_i\sbrac{-\beta_iG''+(1-\beta_iG')^2-G'(1-\beta_iG')}.
\end{align}
\end{subequations}
If the constants $\beta_0,\ldots,\beta_{D-2}$ satisfies
\begin{align}
 \beta_0+\sum_{i=1}^{D-2}\beta_i=1, \label{KasnerCondition1Deriving}
\end{align}
and $G$ as given by \Eqref{LifKas_Functions}, Eqs.~\Eqref{LifKas_EE1} and \Eqref{LifKas_EE2} will reduce to equations identical to the equations of motion for the Lifshitz black hole, namely
\begin{align}
 -z (z +D-2)&=\frac{2\Lambda\ell^2-2(D-3)e^2}{D-2},\\
 -(z +D-2)&=\frac{2\Lambda\ell^2+2e^2}{D-2}.
\end{align}
These are easily solved to give
\begin{align}
 \ell^2=-\frac{(z +D-2)(z +D-3)}{2\Lambda},\quad e=\pm\sqrt{\half(z -1)(z +D-2)}. \label{LifKas_Constants}
\end{align}
If we further assume, for some constant $K$ that
\begin{align}
 \beta_0^2+\sum_{i=1}^{D-2}\beta_i^2=1-K, \label{SecondKasner2Deriving}
\end{align}
and using \Eqref{KasnerCondition1Deriving} and \Eqref{LifKas_EE1} to eliminate $\sum_i\beta_i$, $\ell$, and $e$, Eq.~\Eqref{LifKas_EE3} becomes 
\begin{align}
 -\expo{-2G}\big[-G''+z ^2+D-2-&(3z +D-2)G'+2G'^2+2(1-\beta_0)(z -1)G'+K G'^2\big]\nonumber\\
   &=\frac{2\Lambda\ell^2-2(D-3)e^2}{D-2}+\expo{-2G}\brac{\varphi'^2+\psi'^2}.
\end{align}
Using \Eqref{LifKas_Scalars} and \Eqref{LifKas_Constants}, this equation reduces to 
\begin{align}
 \brac{K-\frac{(z-1)(1-\beta_0)^2}{D-2}-\eta^2}G'^2&=0.
\end{align}
In choosing $K$ so that this equation is satisfied, Eq.~\Eqref{SecondKasner2Deriving} becomes
\begin{align}
 \beta_0^2+\sum_i\beta_i^2=1-\frac{(z-1)(1-\beta_0)^2}{D-2}-\eta^2.
\end{align}
Our metric and fields are now completely determined. Collecting the results together, the solution is 
\begin{align}
 \dif s^2&=\ell^2\brac{-\expo{2z \rho-2\beta_0G}\dif t^2+\sum_{i=1}^{D-2}\expo{2\rho-2\beta_iG}\brac{\dif x^i}^2+\expo{2G}\dif\rho^2},\nonumber\\
 A&=\sqrt{\frac{\ell^2(z -1)}{2(z +D-2)}}\expo{(z +D-2)\rho}\,\dif t,\quad G=-\half\ln\brac{1-c\expo{-(z +D-2)\rho}},\nonumber\\
 \varphi&=\sqrt{(z -1)(D-2)}\rho-(1-\beta_0)\sqrt{\frac{z -1}{D-2}}G,\quad \psi=\eta G,\nonumber\\
 \ell&=-\frac{(z +D-2)(z +D-3)}{2\Lambda},\quad\alpha=\sqrt{\frac{D-2}{z -1}},\nonumber\\
 \beta_0&+\sum_{i=1}^{D-2}\beta_i=1,\quad \beta_0^2+\sum_{i=1}^{D-2}\beta_i^2=1-\frac{(z -1)(1-\beta_0)^2}{D-2}-\eta^2.
\end{align}
Upon transforming to Poincar\'{e} coordinates $\rho=-\ln u$ and redefining $c=u_0^{-\brac{z+D-2}}$, we obtain the metric in the form written in \Eqref{LifKas_Solution}.

\bibliographystyle{lifkas}

\bibliography{lifkas}

\end{document}